\documentclass[article,notitle]{jdssv}
\AtBeginDocument{\maketitle}

\usepackage{thumbpdf,lmodern}
\usepackage{amsmath}
\usepackage{bm}
\usepackage{booktabs}

\usepackage[utf8]{inputenc}

\author{
Stuart Lee\\Monash University \And Ursula Laa\\Monash University \And Dianne Cook\\Monash University
}
\title{Casting Multiple Shadows: High-Dimensional Interactive Data Visualisation with Tours and Embeddings}

\Plainauthor{Stuart Lee, Ursula Laa, Dianne Cook}
\Plaintitle{Casting Multiple Shadows: High-Dimensional Interactive Data Visualisation with Tours and Embeddings}
\Shorttitle{\pkg{liminal}: High-Dimensional Interactive Data Visualisation}

\Abstract{
Non-linear dimensionality reduction (NLDR) methods such as t-distributed stochastic neighbour embedding (t-SNE) are ubiquitous in the natural sciences, however, the appropriate use of these methods is difficult because of their complex parameterisations; analysts must make trade-offs in order to identify structure in the visualisation of an NLDR technique. We present visual diagnostics for the pragmatic usage of NLDR methods by combining them with a technique called the tour. A tour is a sequence of interpolated linear projections of multivariate data onto a lower dimensional space. The sequence is displayed as a dynamic visualisation, allowing a user to see the shadows the high-dimensional data casts in a lower dimensional view. By linking the tour to an NLDR view, we can preserve global structure and through user interactions like linked brushing observe where the NLDR view may be misleading. We display several case studies from both simulations and single cell transcriptomics, that shows our approach is useful for cluster orientation tasks. The implementation of our framework is available as an \proglang{R} package called \pkg{liminal} available at \url{https://github.com/sa-lee/liminal}.
}

\Plainkeywords{dimensionality reduction, high-dimensional data, interactive graphics, t-SNE, grand tour, R}

\Address{
    Stuart Lee\\
    Monash University\\
    Department of Econometrics and Business Statistics,\\
Monash University\\
  E-mail: \email{stuart.lee1@monash.edu}\\
  
      Ursula Laa\\
    Monash University\\
    Department of Econometrics and Business Statistics,\\
Monash University\\

      Dianne Cook\\
    Monash University\\
    Department of Econometrics and Business Statistics,\\
Monash University\\

  }

\begin{document}

\hypertarget{introduction}{%
\section{Introduction}\label{introduction}}

High dimensional data is increasingly prevalent in the natural sciences and
beyond but presents a challenge to the analyst in terms of data cleaning,
pre-processing and visualisation. Methods to embed data from a high-dimensional
space into a low-dimensional one now form a core step of the data analysis
workflow where they are used to ascertain hidden structure and de-noise data
for downstream analysis .

Choosing an appropriate embedding presents a challenge to the analyst. How does
an analyst know whether the embedding has captured the underlying topology and
geometry of the high dimensional space? The answer depends on the analyst's
workflow. \citet{Brehmer2014-hk} characterised two main workflow steps that an
analyst performs when using embedding techniques: dimension reduction and
cluster orientation. The first relates to dimension reduction achieved by using
an embedding method, here an analyst wants to characterise and map meaning onto
the embedded form, for example identifying batch effects from a high throughput
sequencing experiment, or identifying a gradient or trajectory along the
embedded form like changes in cell development or species abundance \citep{Nguyen2019-yh}. The second relates to using embeddings as part of
a clustering workflow. Here analysts are interested in identifying and naming
clusters and verifying them by either applying known labels or colouring by
variables that are a-priori known to distinguish clusters like the expression of a marker gene to identify a cell type. Once clusters are identified they are used for further analysis to identify what features in the data make them distinguishable. Both of these
workflow steps rely on the embedding being representative of the original high dimensional dataset, and becomes much more difficult when there is no underlying ground truth.

As part of a visualization workflow, it is important to consider the perception
and interpretation of embedding methods as well. \citet{Sedlmair2013-pn} showed that
scatter plots were mostly sufficient for detecting class separation, however
they also noted that often multiple embeddings were required. For the task of
cluster identification, \citet{Lewis2012-ai} showed experimentally that novice users
of non-linear embedding techniques were more likely to consider clusters of
points on a scatter plot to be the result of a spurious embedding compared
to advanced users who were aware of the inner workings of the embedding
algorithm.

There is no one-size fits all: finding an appropriate embedding for a given
dataset is a difficult and a somewhat poorly defined problem. For non-linear
methods, there are a lot of parameters to explore that can have an effect on
the resulting visualisation and interpretation.
While there has been much work on the algorithmic details of
embedding methods, there are relatively few tools designed to assist users
to interact with these techniques: when is an embedding sufficient for the
task at hand? Several
interactive interfaces have been proposed for evaluating or using embedding
techniques. \citet{Buja2008-fn} used tours to guide analysts during the optimisation
of multidimensional scaling methods by extending their interactive visualisation
software called \pkg{XGobi} and \pkg{GGobi} into a new tool called \pkg{GGvis}
\citep{Swayne1998-uq, Swayne2003-qd, Swayne2004-ua}.
Their interface allows the analyst to dynamically modify and check whether
an MDS configuration has preserved the locality and closeness
of points between the configuration and the original data.
\citet{Ovchinnikova2020-sy} created the \pkg{Sleepwalk} interface for checking
non-linear embeddings in single cell RNA-seq data. It provides a click and
highlight visualisation for colouring points in an embedding according to an
estimated pairwise distance in the original high-dimensional space.
Similarly, the \pkg{TensorFlow} embedding projector is a web interface to
running some non-liner embedding methods live in the browser and
provides interactions to colour points, and select nearest neighbours
\citep{Smilkov2016-hp}. Finally, the work by \citet{Pezzotti2017-cz} provides a user
guided and modified form of the t-SNE algorithm, that allows users to modify
optimisation parameters in real-time.

A complementary approach for visualizing structure in high dimensional data is
the tour. A tour is a sequence of projections of a high dimensional dataset
onto a low-dimensional basis matrix, and is represented as an
animated visualization \citep{Asimov1985-vp, Buja1986-zr}. Given the dynamic
nature of the tour, user interaction is important for controlling and
exploring the visualisation: the tour has been used previously by
\citet{Wickham2015-cx} for exploring statistical model fits and by \citet{Buja1996-fk} for
exploring the space of factorial experimental designs.

The approach used in this paper is to augment the results of an NLDR method with the tour with our \proglang{R} package called \pkg{liminal}. Interfaces for evaluating
embeddings require interaction but should also be able to be incorporated into
an analysts workflow. Here we implement a more pragmatic workflow by
incorporating interactive graphics and tours with embeddings that allows users
to see a global overview of their high dimensional data and assists them with cluster orientation tasks.

The rest of the paper is organised as follows. The next section provides
background on dimension reduction methods, including an overview of the tour.
Then we describe the visual design of \pkg{liminal}, followed by implementation
details. Next we provide case studies that show how our interface assists
in using embedding methods. Finally, we describe the insights gained by using
\pkg{liminal} and plans for extensions to the software.

\hypertarget{overview-of-dimension-reduction}{%
\section{Overview of Dimension Reduction}\label{overview-of-dimension-reduction}}

To begin we suppose the data is in the form of a rectangular
matrix of real numbers, \(X = [\mathbf{x_1}, \dots, \mathbf{x_n}]\), where \(n\) is
the number of observations in \(p\) dimensions. The purpose of any DR algorithm
is to find a low-dimensional representation of the data,
\(Y = [\mathbf{y_1}, \dots, \mathbf{y_n}]\), such that \(Y\) is an \(n \times d\)
matrix where \(d \ll p\). The hope of the analyst is that the DR procedure to
produces \(Y\) will remove noise in the original dataset while retaining any
latent structure.

DR methods can be classified into two broad classes: linear and non-linear
methods. Linear methods perform a linear transformation of the data, that is,
\(Y\) is a linear transformation of \(X\). One example is principal components
analysis (PCA) which performs an eigendecomposition of the estimated sample
covariance matrix. The eigenvalues are sorted in decreasing order and represent
the variance explained by each component (eigenvector).
A common approach to deciding on the number of principal components to retain is
to plot the proportion of variance explained by each component and choose a
cut-off.
When working with linear transformations we often need more than two dimensions
to capture the latent structure. In this case we can use tour methods
\citep{Asimov1985-vp, Buja1986-zr} to show interpolated sequences of projections,
providing the viewer with intuition about structure in more than two
dimensions, as described below.

For non-linear methods \(Y\) is generated via a pre-processed form of the input
\(X\) such as the \(k\)-nearest neighbours graph or via a kernel transformation.
Multidimensional scaling (MDS) is a class of DR methods that aims to construct
an embedding \(Y\) such that the pair-wise distances (inner products) in \(Y\)
approximate the pair-wise distances (inner products) in \(X\)
\citep{Torgerson1952-am, Kruskal1964-cz}. There are many
variants of MDS, such as non-metric scaling which amounts to replacing
distances with ranks instead \citep{Kruskal1964-cw}.
A related technique is Isomap which uses a \(k\)-nearest neighbour graph
to estimate the pair-wise geodesic distance of points in \(X\) then uses classical
MDS to construct \(Y\) \citep{Silva2003-xw}. Other approaches are based on diffusion
processes such as diffusion maps \citep{Coifman2005-ak}. A recent example of this
approach is the PHATE algorithm \citep{Moon2019-ce}.

A general difficulty with using
non-linear DR methods for exploratory data analysis is selecting and tuning
appropriate parameters. For concreteness here we focus on t-distributed
stochastic neighbour embedding (t-SNE), and we will examine
its underpinning in some detail below \citep{Maaten2008-sk}. Similar considerations hold for related
methods, for example UMAP \citep{McInnes2018-co}.

\hypertarget{introduction-to-t-sne}{%
\subsection{Introduction to t-SNE}\label{introduction-to-t-sne}}

The t-SNE algorithm estimates the pair-wise similarity of points in a high
dimensional space based on their (Euclidean) distances using a Gaussian
distribution. A configuration in the low dimensional embedding space is then
estimated by modelling similarities using a t-distribution with 1 degree of
freedom \citep{Maaten2008-sk}. There are several subtleties
to the to use of the algorithm that are revealed by stepping through its
machinery.

To begin, t-SNE transforms pair-wise distances between \(\mathbf{x_i}\) and
\(\mathbf{x_j}\) to similarities using a Gaussian kernel:

\[ 
p_{i|j} = \frac{\exp(-\lVert \mathbf{x_i - x_j} \rVert ^ 2 /
2\sigma_i^2)}{\sum_{k \ne i}\exp(-\lVert \mathbf{x_j - x_k} \rVert ^ 2 /
2\sigma_i^2)} 
\]

The conditional probabilities are then normalised and symmetrised to form a
joint probability distribution via averaging:

\[ p_{ij} = \frac{p_{i|j} + p_{j|i}}{2n} \]

The variance parameter of the Gaussian kernel is controlled by the analyst
using a fixed value of perplexity for all observations:

\[ \text{perplexity}_i = \exp(-\log(2) \sum_{i \ne j}p_{j|i}\log_2(p_{j|i})) \]

As the perplexity increases, \(\sigma^2_{i}\) increases, until its bounded above
by the number of observations , \(n-1\), in the data, corresponding to
\(\sigma^2_{i} \rightarrow \infty\). This essentially turns t-SNE into a
nearest neighbours algorithm, \(p_{i|j}\) will be close to zero for all
observations that are not in the \(\mathcal{O}(\text{perplexity}_i)\)
neighbourhood graph of the \(i\)th observation \citep{Van_Der_Maaten2014-zn}.

Next, in the target low-dimensional space, \(Y\), pair-wise distances between
\(\mathbf{y_i}\) and \(\mathbf{y_j}\) are modelled as a symmetric probability
distribution using a t-distribution with one degree of freedom (Cauchy kernel):

\[ q_{ij} = \frac{w_{ij}}{Z} \\ \text{where } w_{ij} = \frac{1}{ 1 + \lVert
\mathbf{y_i - y_j} \rVert ^ 2} \text{ and } Z = \sum_{k \ne l} w_{kl}. \]

The resulting embedding \(Y\) is the one that minimizes the Kullback-Leibler
divergence between the probability distributions formed via similarities of
observations in \(X\), \(\mathcal{P}\) and similarities of observations in \(Y\),
\(\mathcal{Q}\):

\[ \mathcal{L(\mathcal{P}, \mathcal{Q})} = \sum_{i \ne j} p_{ij} \log
\frac{p_{ij}}{q_{ij}}\]

Re-writing the loss function in terms of attractive (right) and repulsive
(left) forces we obtain:

\[ \mathcal{L(\mathcal{P}, \mathcal{Q})} = -\sum_{i \ne j} p_{ij}\log w_{ij} +
\log\sum_{i \ne j} w_{ij} \]

Minimising the loss function corresponds to large attractive
forces, that is, the pair-wise distances in \(Y\) are small when there are
non-zero \(p_{ij}\), i.e.~\(x_i\) and \(x_j\) are close together. The repulsive force
should also be small, that is, overall the
pair-wise distances in \(Y\) should be large regardless of the magnitude of the
corresponding distances in \(X\). As a result, clusters that are separate in \(X\)
will be placed far from each other in \(Y\).
This minimisation is done via stochastic gradient decent,
and introduces a number of hyperparameters, for example the number of iterations,
the learning rate, and early exaggeration, a multiplier of the attractive force
used at the beginning of the optimisation.

Taken together, these details reveal the sheer number of decisions that an
analyst must make. How does one choose the perplexity? And the
parameters that control the optimisation of the loss function?
It is a known problem that
t-SNE can have trouble recovering topology and that configurations can be
highly dependent on how the algorithm is initialised and parameterized
\citep{Wattenberg2016-ji, Kobak2019-lm, Melville2020}. If the goal is cluster
orientation a recent theoretical contribution by \citet{Linderman2019-dq} proved that
t-SNE can recover spherical and well separated cluster shapes, and proposed new
approaches for tuning the optimisation parameters. However, the cluster sizes
and their relative orientation from a t-SNE view can be misleading
perceptually, due to the algorithms emphasis on locality.

Another recent method, UMAP, has seen a large rise in popularity (at least in
single cell transcriptomics) \citep{McInnes2018-co}. It is a method that is related to
LargeVis \citep{Tang2016-oz}, and like t-SNE acts on the k-nearest neighbour graph.
Its main differences are that it uses a different cost function (cross entropy)
which is optimized using stochastic gradient descent and defines a different
kernel for similarities in the low dimensional space. Due to its computational
speed it is possible to generate UMAP embeddings in more than three dimensions.
It appears to suffer from the same perceptual issues as t-SNE, however it supposedly preserves global structure better than t-SNE \citep{Coenen2019-to}.

\hypertarget{tours-explore-the-subspace-of-low-dimensional-projections}{%
\subsection{Tours explore the subspace of low dimensional projections}\label{tours-explore-the-subspace-of-low-dimensional-projections}}

The tour is a visualisation technique that is grounded in
mathematical theory, and allows the viewer to ascertain
the shape and global structure of a dataset via inspection of the subspace generated by the set of low-dimensional projections \citep{Asimov1985-vp, Buja1986-zr}.

As with other DR techniques, the tour assumes we have a real data matrix
\(X\) consisting of \(n\) observations in \(p\) dimensions. First, the tour generates
a sequence of \(p \times d\) orthonormal projection matrices (bases)
\(A_t\), where \(d\) is typically 1 or 2. For each pair of
orthonormal bases \(A_{t}\) and \(A_{t+1}\) that are generated, the geodesic
path between them is interpolated to form intermediate frames, giving the
sense of continuous movement from one basis to another. The tour is then the
continuous visualisation of the projected data \(Y_{t} = XA_{t}\), that is the
projection of \(X\) onto \(A_{t}\) as the tour path is interpolated between
successive bases.

A \emph{grand tour} corresponds to choosing new orthonormal
bases at random; allowing a user to ascertain structure via exploring the
subspace of \(d\)-dimensional projections. In practice, we first sphere
our data via principal components to reduce dimensionality of \(X\) prior to
running the tour. Instead of picking projections at
random, a \emph{guided tour} can be used to generate a sequence of `interesting'
projections as quantified by an index function \citep{Cook1995-bi}. While our software,
\pkg{liminal} is able to visualise guided tours, our focus in the case studies
uses the grand tour to see global structure in the data.

\hypertarget{visual-design}{%
\section{Visual Design}\label{visual-design}}

Tours provide a supportive visualisation to NLDR graphics, and
can be easily incorporated into an analysts workflow with our software package,
\pkg{liminal}. Our interface allows analysts to quickly compare
views from embedding methods and see how an embedding method
preserves or alters the geometry of their data. Using multiple concatenated and
linked views with the tour enhances interaction techniques, and allows
analysts to perform cluster orientation tasks via linked highlighting and
brushing \citep{McDonald1982-ew, Becker1987-gd}.
This approach allows our interface to achieve
the three principles for interactive high-dimensional data visualisation outlined
by \citet{Buja1996-fk}: finding gestalt (identifying patterns in visual forms), posing queries, and making comparisons.

\hypertarget{finding-gestalt-focus-and-context}{%
\subsection{Finding Gestalt: focus and context}\label{finding-gestalt-focus-and-context}}

To understand the data structure, we look for Gestalt features such as
(non-)linearities, clusters or outliers. A tour display is
preferred for this task, since it accurately captures the geometry, while NLDR
methods typically introduce distortions.

To investigate latent structure and the shape of a high dimensional dataset in
\pkg{liminal}, a tour can be run without the use of an external embedding.
It is often useful to
first run principal components on the input as an initial dimension reduction
step, and then tour a subset of those components instead, i.e.~by selecting
them from a scree plot. The default tour layout is a scatter plot with an axis
layout displaying the magnitude and direction of each basis vector. Since the
tour is dynamic, it is useful to be able to pause and highlight a
particular view. In addition to pause, play and reset buttons, brushing will
pause the tour path, allowing users to identify `interesting' projections.
The
domain of the axis scales from running a tour is called the half range, and is
computed by rescaling the input data onto \(d\)-dimensional unit cube. We bind
the half range to a mouse wheel event, allowing a user to pan and zoom on the
tour view dynamically. This is useful for peeling back dense clumps of points
to reveal structure.

\hypertarget{posing-queries-multiple-views-many-contexts}{%
\subsection{Posing Queries: multiple views, many contexts}\label{posing-queries-multiple-views-many-contexts}}

The initial visualisation gives an overview of the data structure,
and naturally leads to queries that investigate observed features with
the aim to further characterise them. This is an essential aspect of
our framework, where we use a tour to better characterise
features observed in a NLDR display.

We have combined the tour view in a side by side layout with a scatter plot
view as has been done in previous tour interfaces \pkg{XGobi} and \pkg{DataViewer}
\citep{Buja1986-ku, Swayne1998-uq}. These views are
linked; analysts can brush regions or highlight collections of points in either
view. Linked highlighting can be performed when points have been previously
labelled according to some discrete structure, i.e.~cluster labels are
available. This is achieved via the analyst clicking on groups in the legend,
which causes unselected groupings to have their points become less opaque.
Consequently, simple linked highlighting can alleviate a known downfall of
methods such as UMAP or t-SNE: that is distances between clusters are
misleading. By highlighting corresponding clusters in the tour view, the
analyst can see the relationship between clusters, and therefore obtain a more
accurate representation of the topology of their data.

Simple linked brushing is achieved via mouse-click and drag movements. By
default, when brushing occurs in the tour view, the current projection is
paused and corresponding points in the embedding view are highlighted.
Likewise, when brushing occurs in the embedding view, corresponding points in
the tour view are highlighted. In this case, an analyst can use brushing for
manually identifying clusters and verifying cluster locations and shapes:
brushing in the embedding view gives analysts a sense of the shape and
proximity of cluster in high-dimensional space.

\hypertarget{making-comparisons-revising-embeddings}{%
\subsection{Making comparisons: revising embeddings}\label{making-comparisons-revising-embeddings}}

Combining multiple views of a single data set allows the analyst
to make meaningful comparisons. In \pkg{liminal} the embedding
and tour views are arranged side-by-side for direct cross-checks.

As mentioned previously, when using any DR method, we are assuming the
embedding is representative of the high-dimensional dataset it was computed
from. Defining what it means for embeddings to be `representative` or 'faithful'
to high-dimensional data is ill-posed and depends on the underlying task an
analyst is trying to achieve. At the very minimum, we are interested in
distortions and diffusions of the high-dimensional data. Distortions occur when
points that are near each other in the embedding view are far from each other
in the original dataset. This implies that the embedding is not continuous.
Diffusions occur when points are far from each other in the embedding view are
near in the original data. Whether, points are near, or far is reliant on the
distance metric used; distortions and diffusions can be thought of as the
preservation of distances or the nearest neighbours graphs between the
high-dimensional space and the embedding space. As distances can be noisy in
high-dimensions, ranks can be used instead as has been proposed by \citet{Lee2009-zb}.
Identifying distortions and diffusions allows an analyst to investigate the
quality of their embedding and revise them iteratively.

These checks are done visually using our side-by-side tour and embedding views.
In the simplest case, a local continuity check can be assessed via one to
one linked brushing from the embedding to the tour view. Similarly, diffusions
are identified from linked brushing on the tour view, highlighting points
in the embedding view.

\hypertarget{software-infrastructure}{%
\section{Software Infrastructure}\label{software-infrastructure}}

We have implemented the above design as an open source \proglang{R} package called
\pkg{liminal} \citep{r-liminal}. The package allows analysts to construct concatenated
visualisations, drawn with the \pkg{Vega-Lite} grammar of interactive graphics via
the \pkg{vegawidget} package \citep{Satyanarayan2017-gs, Lyttle2020-hp}. It provides
an interface for constructing linked and stand alone interfaces for
manipulating tour paths via the \pkg{shiny} and \pkg{tourr} packages
\citep{Chang2020-bq, Wickham2011-st}.

\hypertarget{tours-as-a-streaming-data-problem}{%
\subsection{Tours as a streaming data problem}\label{tours-as-a-streaming-data-problem}}

The process of generating successive bases and interpolating between them to\\
construct intermediary frames, means the tour is a dynamic visualisation
technique. Generally, the user would set \(d=2\) and the tour is visualised as
an animated scatter plot. This process of constructing bases and intermediate
frames and visualising the resulting projection is akin to making a ``flip book''
animation. Like with a flip book, an interface to the tour requires the ability
to interact and modify it in real time. The user interface generated in \pkg{liminal}
allows a user to play, pause, and reset the tour animation,
panning and zooming to modify the scales of the plot to provide context and
click events to highlight groups of points if a labelling variable has
been placed on the legend.

These interactions are enabled by treating the basis generation as a reactive
stream. Instead of realising the entire sequence, which limits the animation
to have a discrete number of frames, new bases and their intermediate frames
are generated dynamically via pushing the current projection to the
visualisation interface. The interface listens to events like pressing a button
or mouse-dragging and reacts by pausing the stream. This process allows the
user to manipulate the tour in real time rather than having to fix the number
of bases ahead of time. Additionally, once the user has identified an
interesting projection or is done with the tour, the interface will return the
current basis for use downstream.

\hypertarget{linking-and-highlighting-views-via-interactions}{%
\subsection{Linking and highlighting views via interactions}\label{linking-and-highlighting-views-via-interactions}}

The embedding and tour views are linked together via rectangular brushes;
when a brush is active, points will be highlighted in the adjacent view.
Because the tour is dynamic, brush events that become active will pause the
animation, so that a user can interrogate the current view. By default,
brushing on the embedding view will produce a one-to-one linking with the tour
view. For interpreting specific combinations of clusters, the multiple
guides on the legend can be selected in order to see their relative orientations.
The interface is constructed as a \pkg{shiny} gadget specifically designed for
interactive data analysis. Selections such as brushing regions and the current
tour path are returned after the user clicks done on the interface and become
available for further investigation.

\hypertarget{case-studies}{%
\section{Case Studies}\label{case-studies}}

The next section steps through case studies of our approach using simulations
and an application to single cell RNA-seq data.

The first three case studies use simulations where the cluster structure and
geometry of the underlying data is known. We start with a simple example
where we generated spherical clusters that are embedded well by t-SNE. Then
we move onto more complex examples where the tour provides insight, such
as clusters that have substructure and where there is more complex geometry
in the data.

In the final case study, we apply our approach to clustering
the mouse retina data from \citet{Macosko2015-ot}, and apply the tour
to the process of verifying marker genes that separate clusters.

We \emph{strongly} recommend viewing the linked videos for each case study
while reading. Links to the videos are available in table \ref{tab:vimeo}
and in the figures for each case study. The videos presented show the visual
appearance of the \pkg{liminal} interface, and how we can interact with the tour
via the controls previously described. If you are unable to view the videos,
the figures in each case study consist of screenshots that summarise what is
learned from combining the tour and an embedding view.

\hypertarget{case-study-1-exploring-spherical-gaussian-clusters}{%
\subsection{Case Study 1: Exploring spherical Gaussian clusters}\label{case-study-1-exploring-spherical-gaussian-clusters}}

To begin we look at simulated datasets that reproduce known facts
about the t-SNE algorithm. Our first data set consists of five spherical 5-\(d\)
Gaussian clusters embedded in 10-\(d\) space, each cluster has the same
covariance matrix. We then computed a t-SNE layout with default settings
using the \pkg{Rtsne} package \citep{Rtsne},
and set up the \pkg{liminal} linked interface with grand tour on the 10-\(d\)
observations.

\begin{figure}

{\centering \includegraphics[width=\textwidth,height=0.75\textheight]{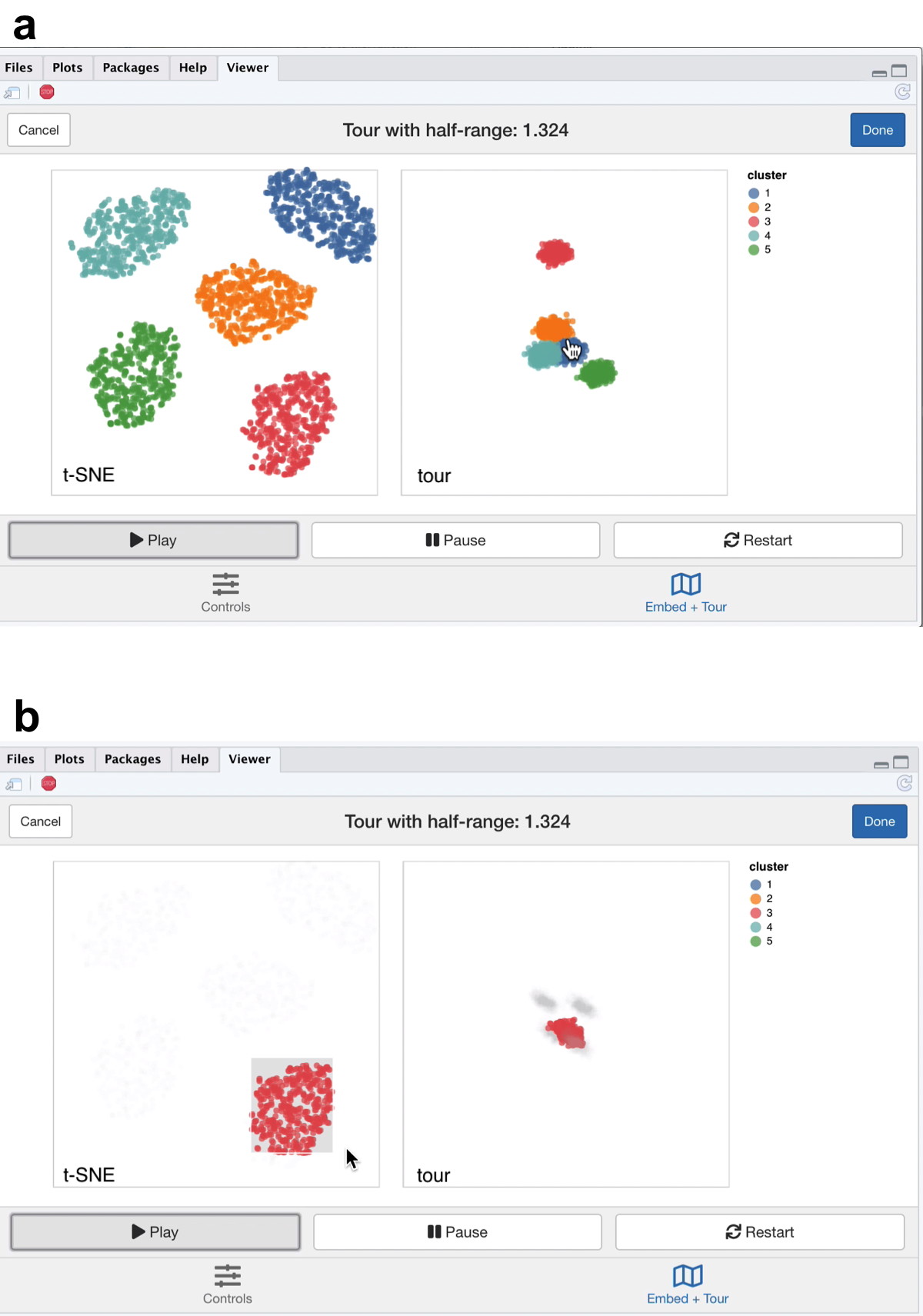} 

}

\caption{Screenshots of the \pkg{liminal} interface applied to well clustered data, a video of the tour animation is available at \url{https://player.vimeo.com/video/439635921}.}\label{fig:gaussian}
\end{figure}

From the video linked in Figure \ref{fig:gaussian}, we learn that t-SNE has
correctly split out each cluster and laid them out in a star like formation.
This agrees with the tour view,
where once we start the animation, the five clusters begin to appear but
generally points are more concentrated in the projection view compared to
the t-SNE layout (Figure \ref{fig:gaussian}a).
This can be seen via brushing the t-SNE view (Figure \ref{fig:gaussian}b).

\hypertarget{case-study-2-exploring-spherical-gaussian-clusters-with-hierarchical-structure}{%
\subsection{Case Study 2: Exploring spherical Gaussian clusters with hierarchical structure}\label{case-study-2-exploring-spherical-gaussian-clusters-with-hierarchical-structure}}

\begin{figure}

{\centering \includegraphics[width=\textwidth,height=0.75\textheight]{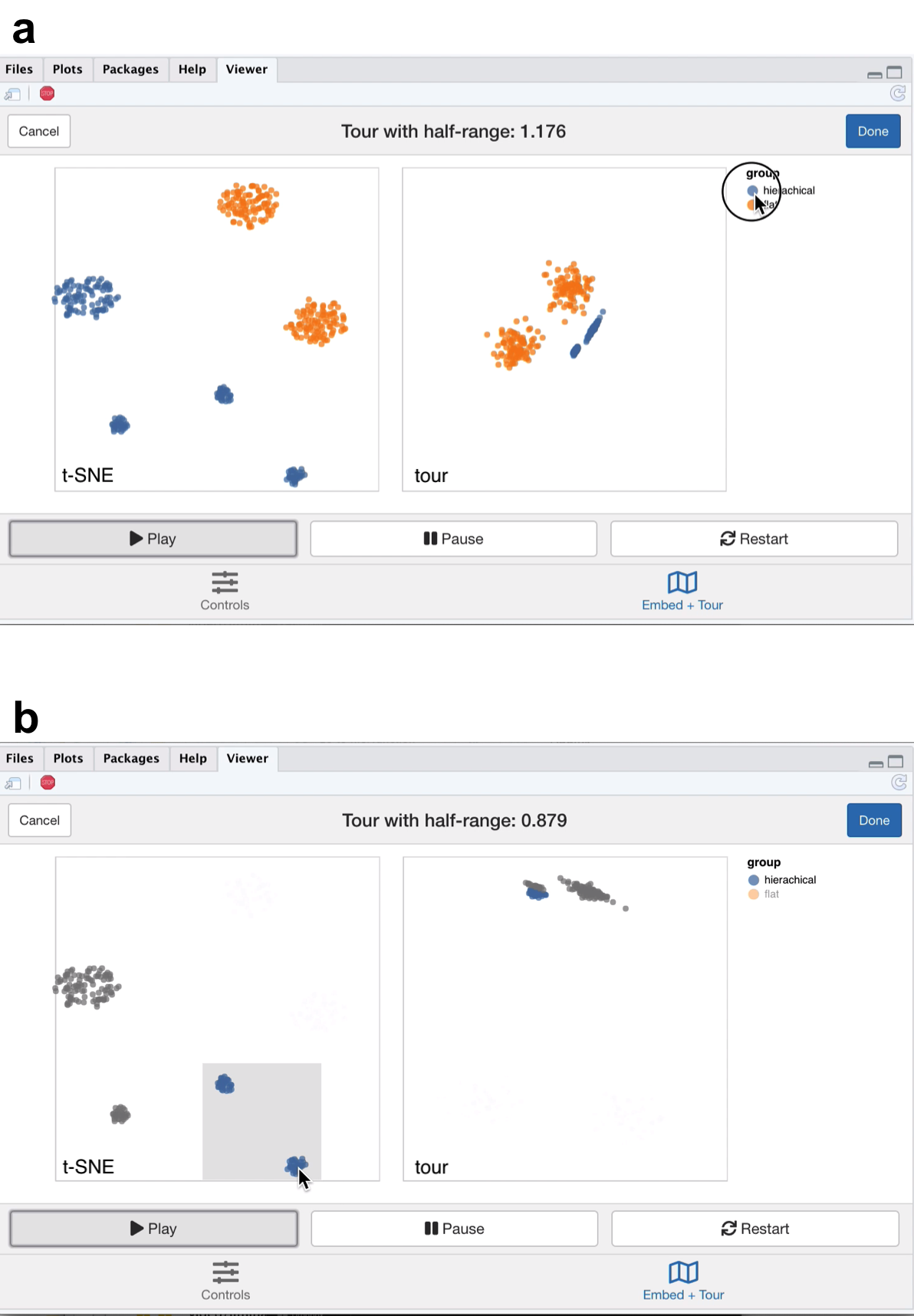} 

}

\caption{Screenshots of the \pkg{liminal} interface applied to sub-clustered data, a video of the tour animation is available at \url{https://player.vimeo.com/video/439635905}.}\label{fig:hierarchical}
\end{figure}

Next we view Gaussian clusters from the
\emph{Multi Challenge Dataset}, a benchmark simulation data set for clustering tasks
\citep{Rauber2009-vh}. This dataset has two Gaussian clusters with equal covariance
embedded in 10-\(d\), and a third cluster
with hierarchical structure. This cluster has two 3-\(d\) clusters embedded in
10-\(d\), where the second cluster is subdivided into three smaller clusters,
that are each equidistant from each other and have the same covariance structure.
From the video linked in Figure \ref{fig:hierarchical}, we see
that t-SNE has correctly identified the sub-clusters. However, their relative
locations to each other is distorted, with the orange and blue groups being
far from each other in the tour view (Figure \ref{fig:hierarchical}a). We see
in this case that is difficult
to see the sub-clusters in the tour view, however, once we zoom and highlight
they become more apparent (Figure \ref{fig:hierarchical}b). When we brush
the sub-clusters in the t-SNE, their
relative placement is again exaggerated, with the tour showing that they
are indeed much closer than the impression the t-SNE view gives.

\hypertarget{case-study-3-exploring-data-with-piecewise-linear-structure}{%
\subsection{Case Study 3: Exploring data with piecewise linear structure}\label{case-study-3-exploring-data-with-piecewise-linear-structure}}

\begin{figure}

{\centering \includegraphics[width=\textwidth]{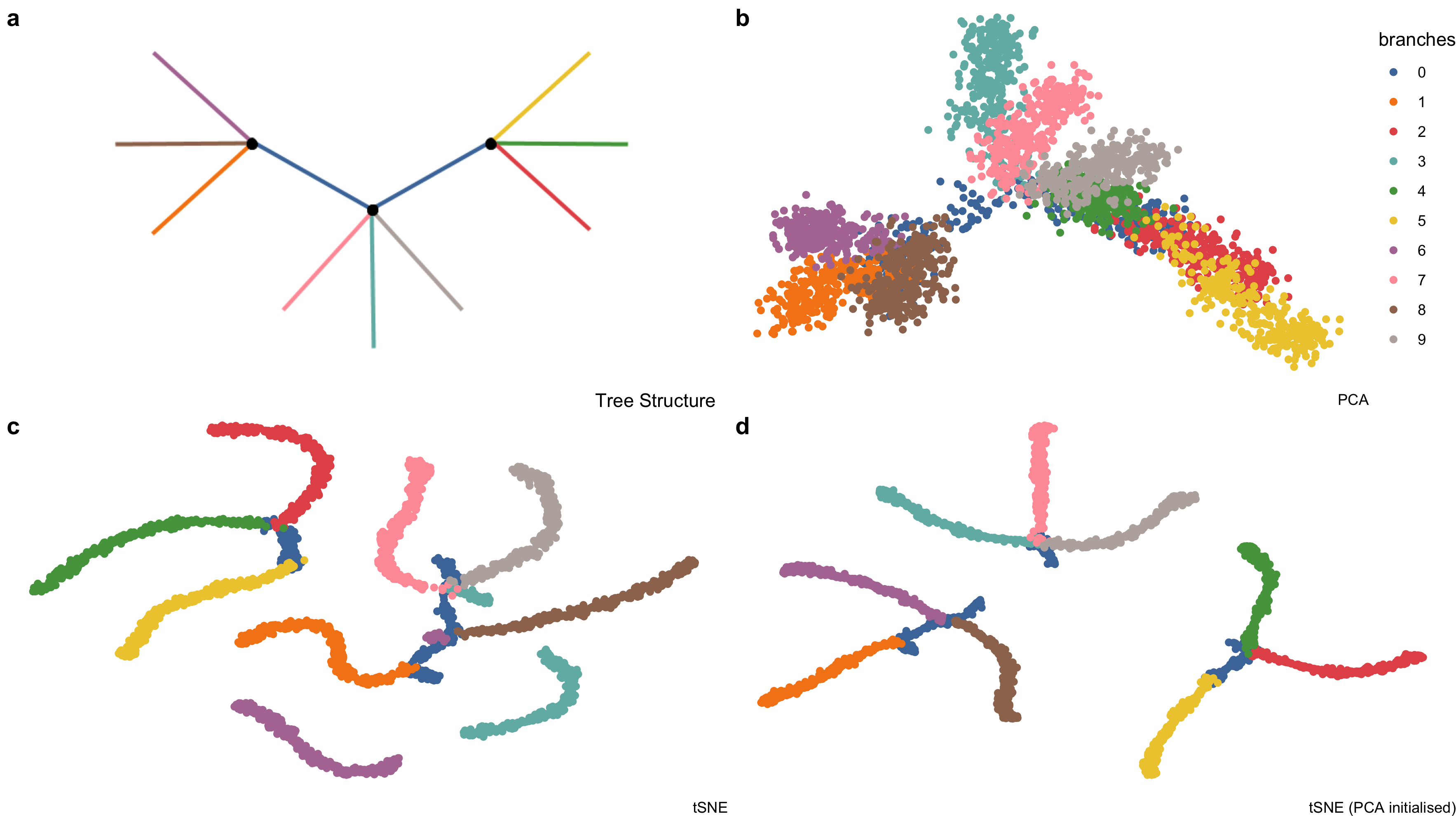} 

}

\caption{Example high-dimensional tree shaped data, \(n = 3000\) and \(p = 100\). \emph{(a)} The true data lies on a 2-\(d\) tree consisting of ten branches. This data is available in the \pkg{phateR} package and is simulated via diffusion-limited aggregation (a random walk along the branches of the tree) with Gaussian noise added \citep{Moon2019-ce}. \emph{(b)} The first two principal components, which form the initial projection for the tour, note that the backbone of the tree is obscured by this view. \emph{(c)} The default t-SNE view breaks the global structure of the tree. \emph{(d)} Altering t-SNE using the first two principal components as the starting coordinates for the embedding, results in clustering the tree at its branching points.}\label{fig:fake-trees}
\end{figure}

Next we explore some simulated noisy tree structured data (Figure
\ref{fig:fake-trees}). Our interest here is how t-SNE visualisations break
topology of the data, and then seeing if we can resolve this by tweaking the
default parameters with reference to the global view of the data set.
This simulation aims to mimic branching trajectories of cell differentiation:
if there were only mature cells, we would just see the tips of the branches
which have a hierarchical pattern of clustering.

First, we apply principal components and restrict the results down to the first
twelve principal components (which makes up approximately 70\% of the variance
explained in the data) to use with the grand tour.

\begin{figure}

{\centering \includegraphics[width=\textwidth,height=0.75\textheight]{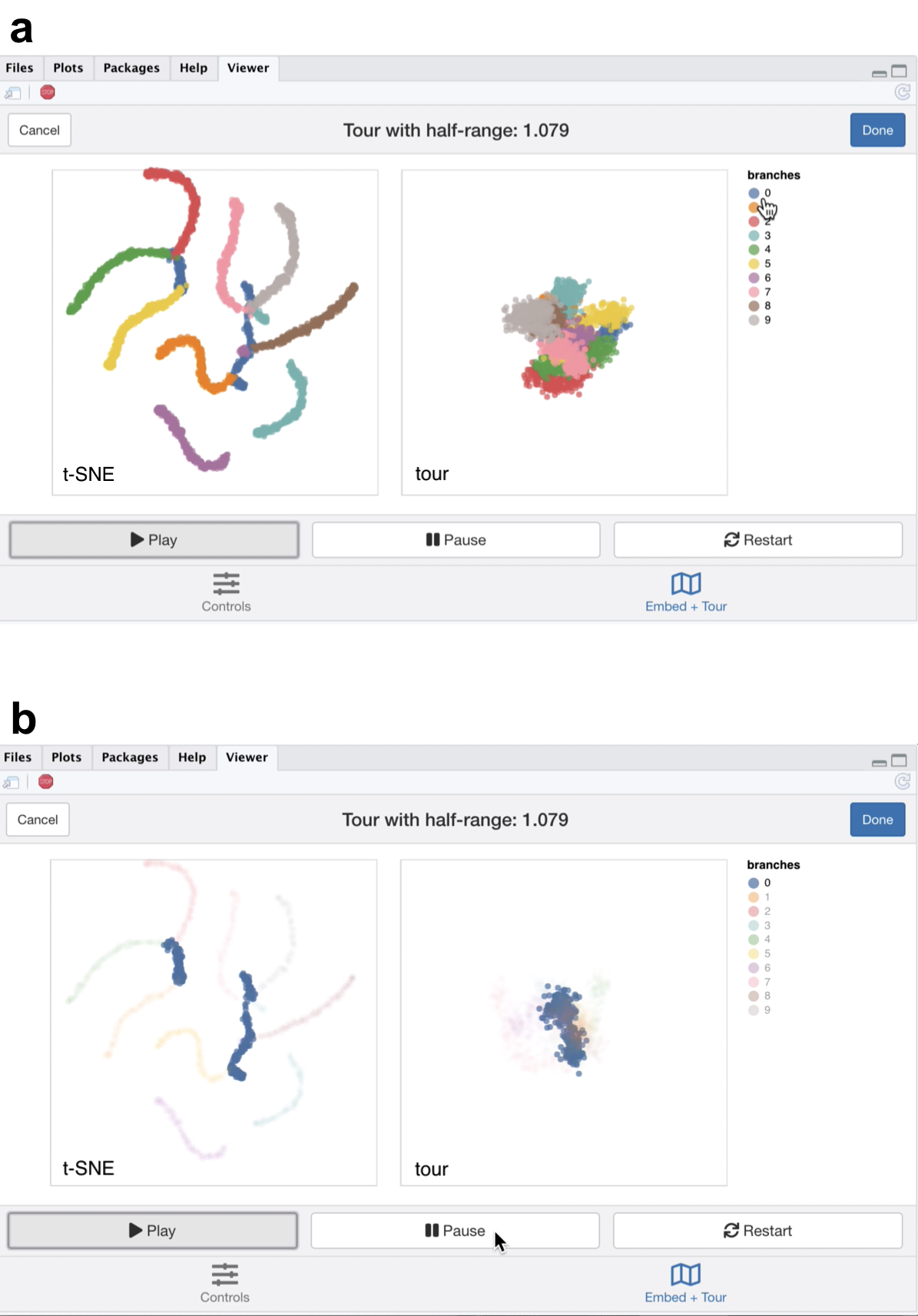} 

}

\caption{Screenshots of the \pkg{liminal} interface applied to tree structured data, a video of the tour animation is available at \url{https://player.vimeo.com/video/439635892}.}\label{fig:trees-01}
\end{figure}

Moreover, we run t-SNE using the default arguments on the complete data (this
keeps the first 50 PCs, sets the perplexity to equal 30 and performs random
initialisation). We then create a linked
tour with t-SNE layout with \pkg{liminal} as shown in Figure \ref{fig:trees-01}.

From the linked video, we see that the t-SNE view has been unable to
recapitulate the topology of the tree - the backbone (blue) branch has been
split into three fragments (Figure \ref{fig:trees-01}a). We can see
this immediately via the linked highlighting over both
plots. If we click on the legend for the zero branch, the blue coloured points
on each view are highlighted and the remaining points are made transparent.
From here it becomes apparent from the tour view that the blue branch forms
the backbone of the tree and is connected to all other branches. From the video
it is easy to see that cluster sizes formed via t-SNE can be misleading; from the tour
view there is a lot of noise along the branches, while this does not appear to
be the case for the t-SNE result (Figure \ref{fig:trees-01}b).

\begin{figure}

{\centering \includegraphics[width=\textwidth]{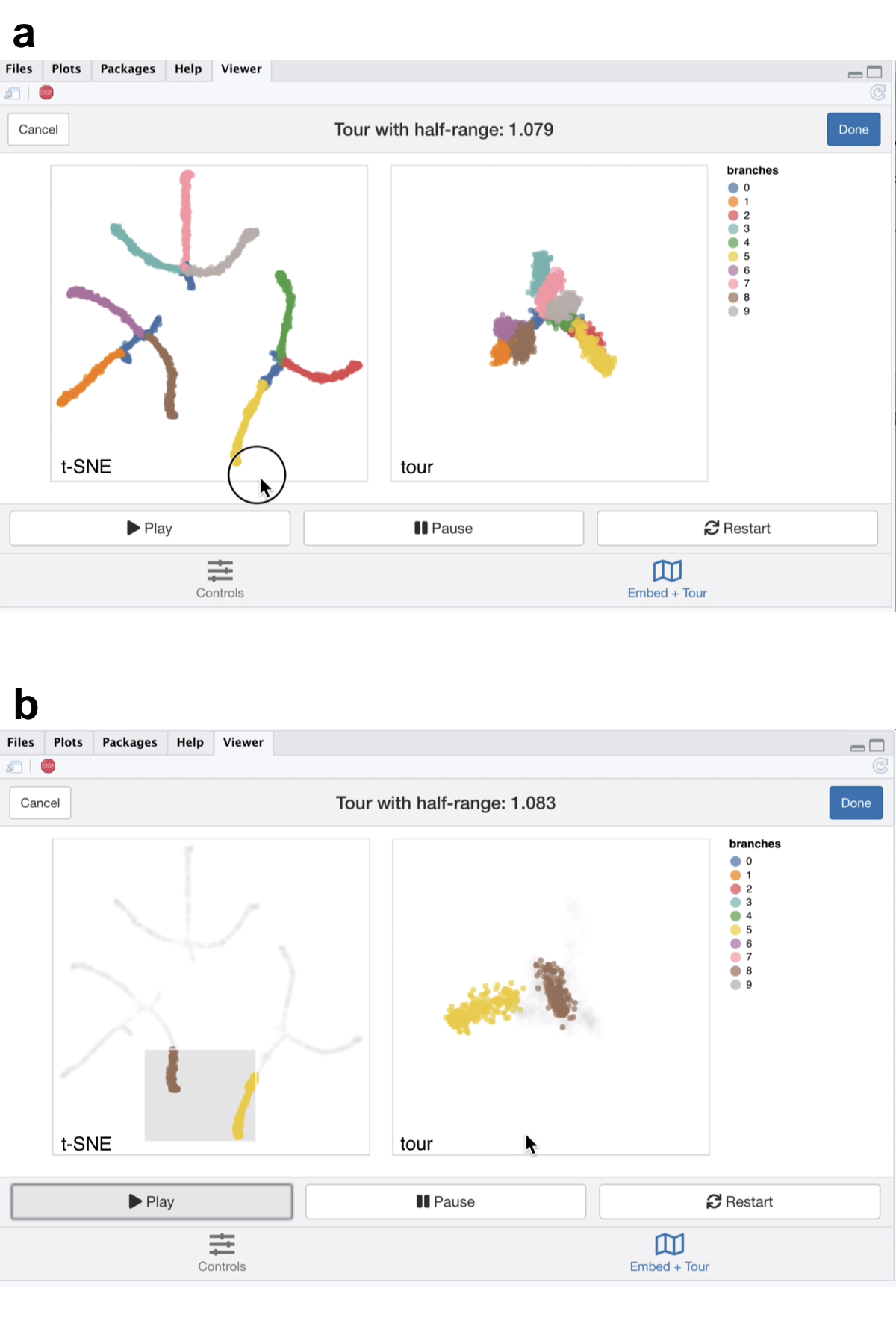} 

}

\caption{Screenshots of the \pkg{liminal} interface applied to tree structured data, a video of the tour animation is available at \url{https://player.vimeo.com/video/439635863}.}\label{fig:trees-02}
\end{figure}

From the first view, we modify the inputs to the t-SNE view, to try and produce
a better trade-off between local structure and retain the topology of the data.
We keep every parameter the same except that we initialise \(Y\) with the first
two PCs (scaled to have standard deviation 1e-4) instead of the default random
initialisation and increase the perplexity from 30 to 100. We then combine
these results with our tour view as displayed in the linked video in the caption
of Figure \ref{fig:trees-02}.

The video linked in Figure \ref{fig:trees-02} shows that this selection of
parameters results in the
tips of the branches (the three black dots in Figure \ref{fig:fake-trees}a)
being split into three clusters representing the terminal branches of the tree.
However, there are perceptual issues following the placement of the three
groupings on the t-SNE view that become apparent via simple linked brushing.
If we brush the
tips of the yellow and brown branches (which appear to be close to each other
on the t-SNE view), we immediately see the placement is distorted in the t-SNE
view, and in the tour view these tips are at opposite ends of the tree
(Figure \ref{fig:trees-02}b).
Although, this is a known issue of the t-SNE algorithm, we can easily identify
it via simple interactivity without knowing the inner workings of the method.

\hypertarget{case-study-4-clustering-single-cell-rna-seq-data}{%
\subsection{Case Study 4: Clustering single cell RNA-seq data}\label{case-study-4-clustering-single-cell-rna-seq-data}}

A common analysis task in single cell studies is performing clustering to
identify groupings of cells with similar expression profiles. Analysts
in this area generally use non linear DR methods for verification and identification
of clusters and developmental trajectories (i.e., case study 1). For clustering
workflows the primary task is verify the existence of clusters and then begin to
identify the clusters as cell types using the expression of ``known''
marker genes. Here a `faithful' a embedding should ideally preserve the topology
of the data; cells that correspond to a cell type should lie in the same
neighbourhood in high-dimensional space. In this case study we use our
linked brushing approaches to look at neighbourhood preservation
and look at marker genes through the lens of the tour. The data we have
selected for this case study has features similar to those found in case studies
2 and 3.

First, we downloaded the raw mouse retinal single cell RNA-seq data from
\citet{Macosko2015-ot} using the \textbf{scRNAseq} Bioconductor package \citep{scRNAseq-d}. We
have followed a standard workflow for pre-processing and normalizing this data
(described by \citet{Amezquita2020-at}): we performed QC using the \textbf{scater} package
by removing cells with high proportion of mitochondrial gene expression and low
numbers of genes detected, we log-transformed and normalised the expression
values and finally selected highly variable genes (HVGs) using \textbf{scran}
\citep{McCarthy2017, Lun2016}. The top ten percent of HVGs were used to subset
the normalised expression matrix and compute PCA using the first 25 components.
Using the PCs we built a shared nearest neighbours graph (with \(k = 10\)) and
used Louvain clustering to generate clusters \citep{Blondel2008-bx}.

To check and verify the clustering we construct a \pkg{liminal} view.
We tour the first five PCs (approximately 20\% of the variance in expression),
alongside the t-SNE view which was computed from all 25 PCs. We have selected only the first five PCs because there is a large drop in the percentage of variance explained after the fifth component, with each component after contributing less than one percent of variance.
Consequently, increasing the number of PCs to tour would increase the dimensionality and volume of the subspace we are touring but without adding any additional signal to the view.

\begin{figure}

{\centering \includegraphics[width=\textwidth,height=0.9\textheight]{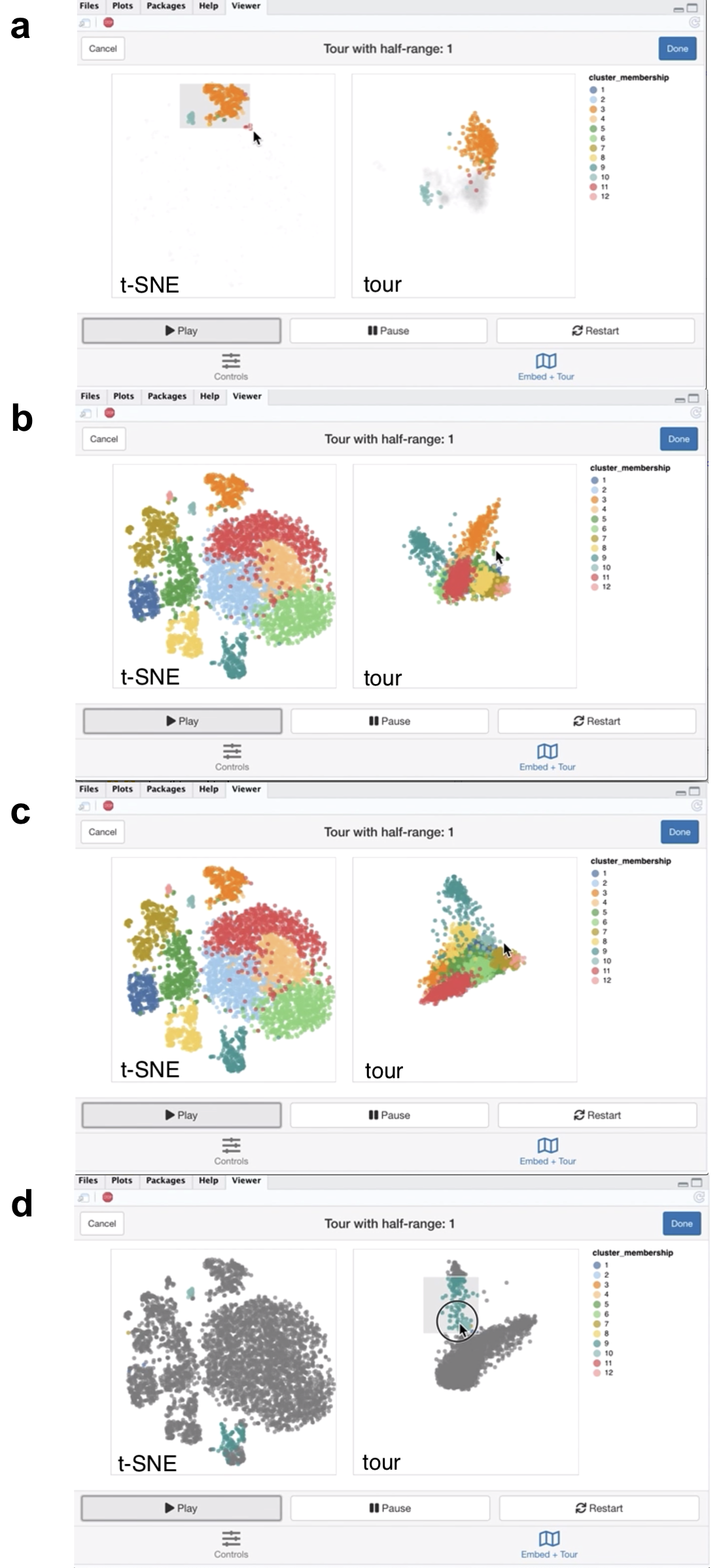} 

}

\caption{Screenshots of the \pkg{liminal} interface applied to single cell data, a video of the tour animation is available at \url{https://player.vimeo.com/video/439635812}.}\label{fig:mouse-01}
\end{figure}

Due to latency of the \pkg{liminal} interface we do a weighted sample of the rows based on
cluster membership, leaving us with approximately 10 per cent of the original
data size - 4,590 cells. Although this is not ideal, it still allows us to
get a sense of the shape of the clusters as seen from the linked video in Figure \ref{fig:mouse-01}. If one was interested in performing more in-depth cluster analysis we recommend an iterative approach of removing large clusters and then re-running the \pkg{liminal} view as a way finding more granular cluster structure.

From the video linked in Figure \ref{fig:mouse-01}, we learn that the embedding has mostly captured the
clusters relative location to each other to their location in high dimensional space, with a notable exception of points in cluster 3 and 10 as shown with
linked brushing (Figure \ref{fig:mouse-01}a). As expected, t-SNE mitigates the crowding problem that is
an issue for tour in this case, where points in clusters 2, 4, 6, and 11 are
clumped together in tour view, but are blown up in the embedding view (Figure \ref{fig:mouse-01}b).
The tour appears to form a tetrahedron like shape, with points lying on
the surface and along the vertices of the tetrahedron in 5-\(d\) PCA space -
a phenomena that has also been observed in \citet{Korem2015-af}
(Figure \ref{fig:mouse-01}c). Brushing on
the tour view, reveals points in cluster 9 that are diffuse in the embedding
view, points in cluster 9 are relatively far away and spread apart from other
clusters in the tour view, but has points placed in cluster 3 and 9 in the embedding
(Figure \ref{fig:mouse-01}d).

Next, we identify marker genes for clusters using one sided Welch t-tests with
a minimum log fold change of one as recommended by \citet{Amezquita2020-at}, which
uses the testing framework from \citet{McCarthy2009-qx}. We select the top 10 marker
genes that are upregulated in cluster 2, which was one of the clumped clusters
when we toured on principal components. Here the tour becomes an alternative
to a standard heatmap view for assessing shared markers; the basis
generation (shown as the biplot on the left view) reflects the relative
weighting of each gene. We run the tour directly on the
log normalised expression values using the same subset as before.

\begin{figure}

{\centering \includegraphics[width=\textwidth,height=0.75\textheight]{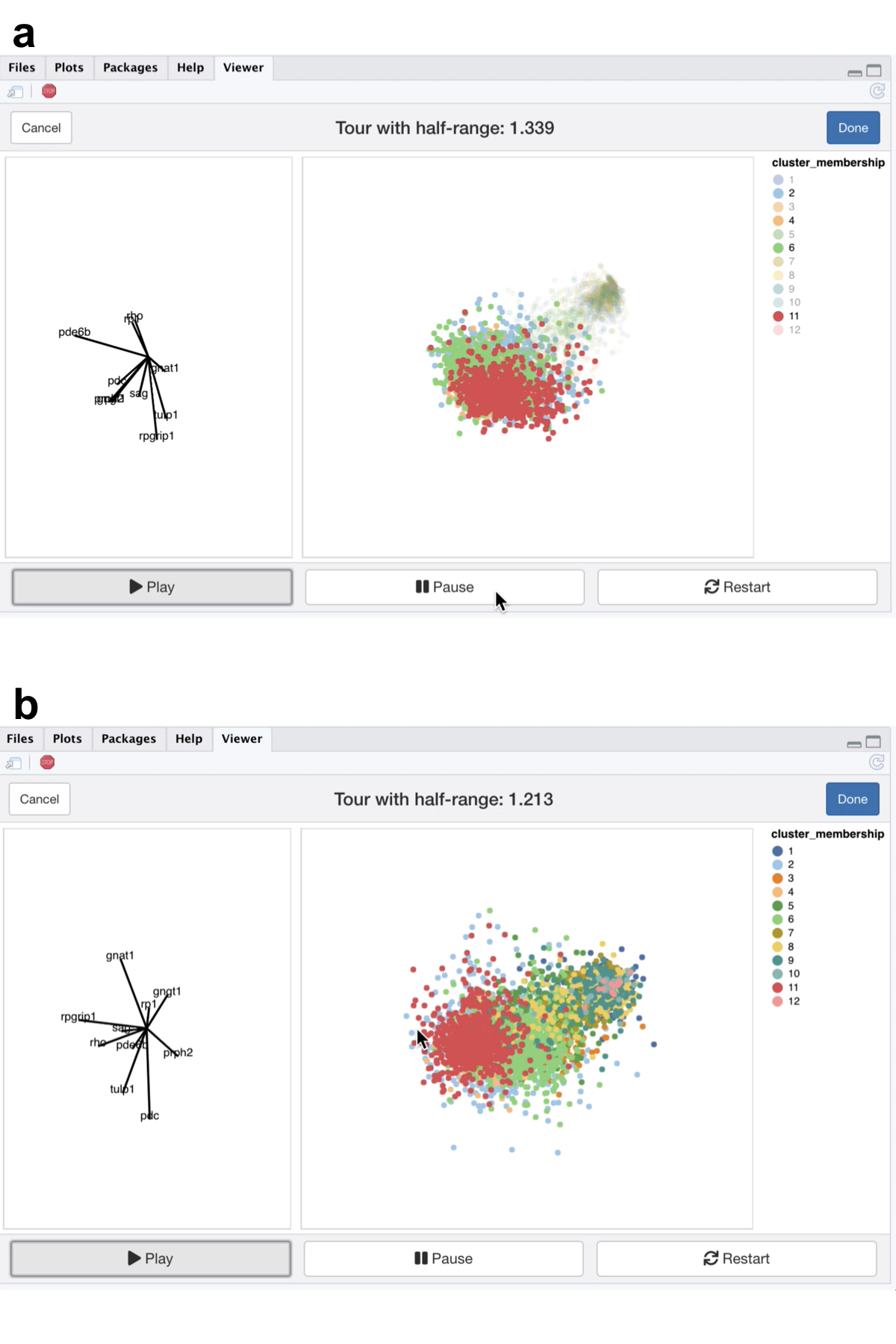} 

}

\caption{Screenshots of the \pkg{liminal} tour applied to a marker gene set, a video of the tour animation is available at \url{https://player.vimeo.com/video/439635843}.}\label{fig:mouse-02}
\end{figure}

From the video linked in Figure \ref{fig:mouse-02}, we see that the
expression of the marker genes, appear to separate
the previously clumped clusters 2, 4, 6, and 11 from the other clusters,
indicating that these genes are expressed in all four clusters (Figure \ref{fig:mouse-02}a). After zooming, we can see a trajectory forming along
the clusters, while the axis view shows that magnitude of expression in
the marker genes is similar across these separated clusters which is
consistent with the results of marker gene analysis (Figure \ref{fig:mouse-02}b).

\hypertarget{discussion}{%
\section{Discussion}\label{discussion}}

We have shown that the use of tours as a tool for interacting with
high dimensional data provides an additional insight for interrogating views
generated from embeddings. The interface we have designed in the \pkg{liminal}
package, allows a user to gain a deeper understanding of an embedding
algorithm, and rectifies perceptual issues associated with NLDR methods via
linked interactions with the tour. As we have shown in the simulation case
studies, the t-SNE method can produce misleading embeddings which can be
detected through the use linked brushing and highlighting. In the case when
the data has a piecewise linear geometry, like the tree simulation, the tour
preserves the shape of the data which can be obscured by the embedding method.

Our framework can also be useful in practice, as displayed
in the fourth case study. The tour when combined with t-SNE allowed us
to identify clusters, while giving us an idea of their orientation to
each other. Moreover, we could visually inspect the separation of clusters
using a tour on marker gene sets. We see our approach as being valuable
to the single cell analyst who wants to make their embeddings more
interpretable.

We have shown in the case studies, that one to one linked brushing can
be used to identify distortions in the embedding, however we would
like extend this to one to many linked brushing, which would allow us to directly interrogate neighbourhood preservation. This form of brushing acts directly on a \(k\)-nearest neighbours (\(k\)-nn) graph computed from a
reference dataset: when a user brushes over a region in the embedding,
all the points that match the graphs edges are selected on the
corresponding tour view. The reference data set for computing
nearest neighbours (for example a distance matrix, or the complete data matrix)
can be independent of the tour or embedding views. In place of
highlighting, one could use opacity or binned colour scales to encode
distances or ranks instead of the neighbouring points. We have begun
implementing this brush in \pkg{liminal}, using the \textbf{FNN} or \textbf{RcppAnnoy}
packages for fast neighbourhood estimation on the server side, however there
are still technicalities that need be resolved \citep{fnn-pkg, rcpp-annoy-pkg}.
Brush composition, such as `and', `or', or `not' brushes, could be used to
further investigate mismatches between the \(k\)-nn graphs estimated from both the embedding and tour views.

There are some limitations
in using the \pkg{liminal} interface for larger datasets. First,
t-SNE avoids the crowding problem, points are separated into distinct regions
on the display canvas. For the tour, points are concentrated in the
centre of the projection and become difficult to see. We have recently proposed a simple non-linear transformation for the tour called a sage tour that aims to fix this problem \citep{Laa2020-vk}.
Second, as \(n\) increases both the embedding view and tour view become harder to read due to over-plotting, while the interactivity and animation become slower as there is more data passing from the server to the client.
For the tasks we have looked at in this paper, where shape and density are important to the analyst, we think that better displays and sub-sampling strategies are more useful than being able to look at every single point on the canvas. We showed in our single cell clustering case study that doing a weighted sample based on cluster membership still allowed us to get a sense of relative cluster orientation, however there are alternative sampling approaches that could be applied, like selecting points close to the cluster centres. Alternative displays via statistical transformations could also mitigate the need to show all of the data. Recent work
by \citet{Laa2020-wr} is a promising area for further investigation, as well as
work from topological statistics \citep{Rieck2017-kk, Genovese2017-iq}.

\hypertarget{acknowledgements}{%
\section*{Acknowledgements}\label{acknowledgements}}
\addcontentsline{toc}{section}{Acknowledgements}

The authors gratefully acknowledge the support of the Australian Research Council.
We would like to thank Dr David Frazier and Dr Paul Harrison for
their feedback on this work. The screenshots and images were compiled using the
\pkg{cowplot} and \pkg{ggplot2} packages \citep{r-cowplot, Wickham2016-gz}.

\hypertarget{supplementary-materials}{%
\section*{Supplementary Materials}\label{supplementary-materials}}
\addcontentsline{toc}{section}{Supplementary Materials}

Code, data, and video for reproducing this paper are available at \url{https://github.com/sa-lee/paper-liminal}. Direct links to videos for viewing
online are available in Table \ref{tab:vimeo}.

\begin{table}

\caption{\label{tab:vimeo}Case Study Videos}
\centering
\begin{tabular}[t]{rll}
\toprule
Case Study & Example & URL\\
\midrule
1 & gaussian & https://player.vimeo.com/video/439635921\\
2 & hierarchical & https://player.vimeo.com/video/439635905\\
3 & trees-01 & https://player.vimeo.com/video/439635892\\
3 & trees-02 & https://player.vimeo.com/video/439635863\\
4 & mouse-01 & https://player.vimeo.com/video/439635812\\
\addlinespace
4 & mouse-02 & https://player.vimeo.com/video/439635843\\
\bottomrule
\end{tabular}
\end{table}

\renewcommand\refname{References}
\bibliography{liminal.bib}

\end{document}